\documentclass{aastex61}

\shorttitle{Shallow Transits Deep Learning I.}
\shortauthors{Zucker and Giryes}

\begin{document}

\title{Shallow Transits --- Deep Learning I:\\
Feasibility Study of Deep Learning to Detect Periodic Transits of Exoplanets}

\correspondingauthor{Shay Zucker}
\email{shayz@post.tau.ac.il}

\author[0000-0003-3173-3138]{Shay Zucker}
\affiliation{School of Geosciences, Raymond and Beverly Sackler Faculty of 
Exact Sciences, Tel Aviv University, Tel Aviv, 6997801, Israel}

\author[0000-0002-2830-0297]{Raja Giryes}
\affiliation{School of Electrical Engineering, Iby and Aladar Fleischman 
Faculty of Engineering, Tel Aviv University, Tel Aviv, 6997801, Israel}

\begin{abstract}

Transits of habitable planets around solar-like stars are expected to be shallow, and to have long periods, which means low information content. The current bottleneck in the detection of such transits is caused in large part by the presence of red (correlated) noise in the light curves obtained from the dedicated space telescopes. Based on the groundbreaking results deep learning achieves in many signal and image processing applications, we propose to use deep neural networks to solve this problem. We present a feasibility study, in which we applied a convolutional neural network on a simulated training set. The training set comprised light curves received from a hypothetical high-cadence space-based telescope. We simulated the red noise by using Gaussian Processes with a wide variety of hyperparameters. We then tested the network on a completely different test set simulated in the same way. Our study proves that very difficult cases can indeed be detected. Furthermore, we show how detection trends can be studied, and detection biases be quantified. We have also checked the robustness of the neural-network performance against practical artifacts such as outliers and discontinuities, which are known to affect space-based high-cadence light curves. Future work will allow us to use the neural networks to characterize the transit model and identify individual transits. This new approach will certainly be an indispensable tool for the detection of habitable planets in the future planet-detection space missions such as PLATO.

\end{abstract}


\section{Introduction} \label{sec:intro}

Ever since the discovery of HD\,209458\,b \citep{Chaetal2000}, 
it has become common knowledge that given the necessary geometrical
conditions, extrasolar planets ('exoplanets') may partially eclipse
their host stars, causing periodic dimming of the observed stellar
light. This phenomenon is commonly dubbed 'transits', alluding to the
solar transits by Mercury and Venus. Currently, this photometric
phenomenon constitutes the dominant channel to detect exoplanets.

The small physical size of a typical planet compared to that of a
solar like star, combined with the typical short duty cycle of a
transit signal, render this phenomenon quite difficult to detect. The
advent of modern photometric methods and the use of space telescopes
facilitate the detection of Jovian planets, and even terrestrial
planets in some fortuitous circumstances. This is achieved by
increasing the photon collection power, thus reducing the uncorrelated
('white') photon noise.

Under the assumption that the dominant noise is the uncorrelated
photon noise, the algorithms to detect planetary transit signals in stellar
light curves have been tackling the challenge in essentially linear approaches. In one way or
another, most of the algorithms use a least-squares approach to find a
best-fitting transit model to the data, e.g., the BLS
\citep{Kovetal2002}. Usually, those techniques essentially scan the parameter space of the
transit model to search for the best-fitting configuration, if such exists.

\citet{Ponetal2006} were the first to demonstrate how correlated
('red') noise can be detrimental to traditional transit detection
approaches. At that time, when all of the transit surveys were still
ground-based, the main source of correlation in the light curve was
considered to be related to instrumental and telluric effects. When data began to flow
from the dedicated space missions, CoRoT \citep{Auvetal2009} and Kepler
\citep{Boretal2010}, it became clear that stellar variability was a
major source of red noise \citep{Aigetal2009,McQetal2012,Basetal2013},
especially in the most difficult signals: transits of terrestrial exoplanets, orbiting in the habitable zones of
solar like stars. Those transits would be very shallow, with long
periods --- meaning a very small effective signal-to-noise ratio
(SNR).

\citet{Cubetal2017} provide a review of the simplest approaches to
correct for the effects of red noise. However, those approaches are essentially equivalent to merely
inflating the effective uncertainties used in the analysis. This is especially relevant
to model fitting, rather than detection. Other approaches try
to use the large database of the survey to partially filter out the
red noise \citep[e.g.,][]{Tametal2005,Kovetal2005}. The current state-of-the-art approaches to deal with red noise try to use GPs --- Gaussian Processes \citep[e.g.,][]{RasWil2006,Aigetal2016}. The
flexibility of GPs allows the modeling of many kinds of time
correlation while maintaining the inherent randomness of the noise. In
spite of impressive progress in the computational treatment of GPs,
such as the Celerite code \citep{Foretal2017}, they still remain very
demanding computationally. Simultaneously modeling a GP while searching
for transits is still out of reach.

In this paper we examine the feasibility of a new approach ---
using deep learning. Concisely speaking, deep learning is a set of computational heuristics to train highly nonlinear parametric functions structured in a layered form to perform a certain task \citep{Lecetal2015,Sch2015,Gooetal2016}. These functions, known as neural networks, are essentially concatenations of basic units (layers), each comprising a linear operation followed by a simple non-linearity, eventually resulting in intricate highly non-linear functionality. A variant of these networks are convolutional neural networks (CNNs), in which convolutions are used for the linear part of the layers \citep{Lecetal1998}. CNNs are widely used with images or periodic signals due to their shift invariance property. 
For the nonlinearity in the layers it is common to use element-wise activation functions such as the sigmoid, hyperbolic tangent and the rectified linear unit (ReLU) \citep{NaiHin2010}, which we employ in this work. To date, deep learning techniques achieve state-of-the-art results in many fields including computer vision, speech processing, machine translation, to name a few \citep{Lecetal2015,Sch2015,Gooetal2016}. In this work we demonstrate the potential of using CNNs to detect planetary transits in high-cadence light curves plagued by red noise.

Section \ref{sec:data} presents the simulated training and testing datasets we used, while Section \ref{sec:convnet} presents the details of the deep learning neural network we applied on those datasets. We present our results in Section \ref{sec:results} and finally discuss their meaning and future implications in Section \ref{sec:discuss}.

\section{Simulated data} \label{sec:data}

We have used simulated data to train the network, and later to test and study its performance. Specifically we generated $10^5$ pure noise light curves, each one with or without a transit signal, amounting to $2\times10^5$ light curves in total. We used the first $2\times83333$ light curves for training, and tested the neural network on the remaining $2\times16667$ light curves.
In the simulation we assumed the data were obtained by a fictitious planet-detection space telescope, with a five-minute cadence, in between Kepler's long-cadence rate of about $30$ minutes \citep{Boretal2010} and the very short $25$-seconds cadence planned for PLATO \citep{Rauetal2014}.  We further assumed the duration of the mission's observing run to be $21.33$ days, which amounts to a convenient number of $6144$ samples.

The fictitious stellar population observed by our mission comprised stars with apparent magnitudes ranging between $V=10$ and $V=16$. Specifically, we drew the magnitudes from a beta distribution  \citep[e.g.,][]{Johetal1995} with coefficients $\alpha=3$ and $\beta=1$, and later linearly transformed the magnitudes to the required interval $[10,16]$. This resulted in a distribution with a preference towards the faint end.

The magnitude directly affects the uncorrelated photon noise, in a way that should depend on the characteristics of the observational apparatus. Inspired by the pre-flight estimates for Kepler \citep{vanCal2009}, we assumed the following relation for the white-noise RMS:
\begin{equation}
A_w =  e^{0.4(V-10)/2} \cdot 35 \,\mu\mathrm{mag}
\end{equation}In order to produce the correlated (red) noise, we have used a GP approach, with a kernel comprising a squared-exponential and a quasi-periodic components \citep[e.g.,][]{Aigetal2016}. Combined with the white-noise component, we got the following expression for the kernel of the noise GP:

\begin{equation}
k(t_i,t_j) = A_s^2 \exp\left[-\left(\frac{t_i-t_j}{\lambda_s}\right)^2\right]+A_q^2 \exp \left[ -\frac{1}{2} \sin^2 \left(\frac{\pi \left(t_i-t_j\right))}{T_q}\right)-\left(\frac{t_i-t_j}{\lambda_q}\right)^2\right] +A_w^2\delta\left(t_i-t_j\right)
\end{equation}
We drew the various hyper-parameters of the kernel from log-uniform distributions over specified intervals detailed in Table \ref{table:hyper}.

\begin{deluxetable}{ccc}
\tablewidth{0pt}
\tablecaption{GP kernel hyper-parameter ranges \label{table:hyper}}
\tablehead{
\colhead{Hyper-parameter} & 
\colhead{Minimum value} & 
\colhead{Maximum value}}
\startdata
$A_s$ & $5\,\mu\,\mathrm{mag}$ & $125\,\mu\,\mathrm{mag}$ \\
$A_q$ & $50\,\mu\,\mathrm{mag}$ & $125\,\mu\,\mathrm{mag}$ \\
$\lambda_s$ &  $1\,\mathrm{min} $ & $10\,\mathrm{h}$ \\
$T_q$ & $10\,\mathrm{h}$ & $500\,\mathrm{h}$ \\
$\lambda_q$ &  $1000\,\mathrm{min} $ & $500\,\mathrm{h}$ \\
\enddata
\end{deluxetable}

We included each red-noise light curve in the data set once as is, and once with a transit signal added. As this study is a feasibility study, we simulated only transits with periods shorter than $6100\,\mathrm{min}$ (about four days), ensuring that at least five transit events would occur during the duration of the observing run. We simulated transits as simple trapezoids, without invoking the full detailed model including limb darkening \citep{ManAgo2002}. We also assumed transits to be strictly periodic, avoiding effects related to multiple planets and transit-timing variations (TTV). We drew the transit phase from a uniform distribution, and we also drew the combined ingress-egress relative duration (as a fraction of the transit duration) from a uniform distribution, between $0.01$ and $0.3$. The two other transit parameters (depth and duration) were drawn from log-uniform distributions as detailed in Table \ref{table:transitparms}.

\begin{deluxetable}{ccc}
\tablewidth{0pt}
\tablecaption{Transit parameter ranges \label{table:transitparms}}
\tablehead{
\colhead{Transit parameter} & 
\colhead{Minimum value} & 
\colhead{Maximum value}}
\startdata
depth & $0.1\,\mathrm{millimag}$ & $1\,\mathrm{millimag}$ \\
duration & $30\,\mathrm{min}$ & $200\,\mathrm{min}$ \\
\enddata
\end{deluxetable}

\section{The Convolutional Neural Network} \label{sec:convnet}

\begin{figure}
\plotone{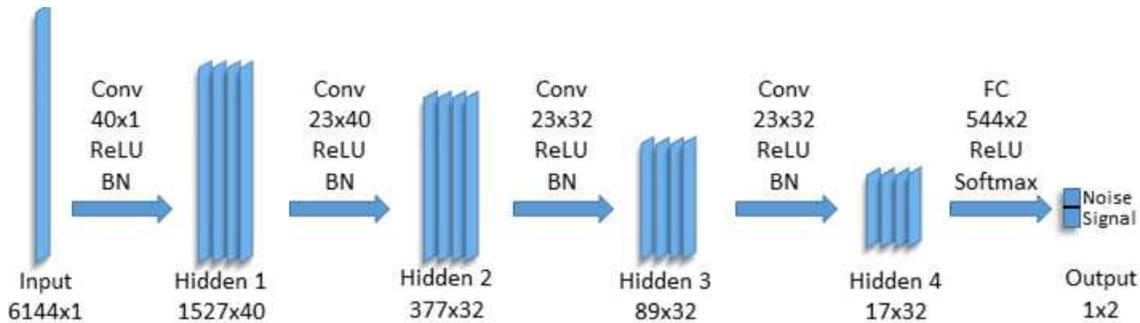}
\caption{Convolutional neural network architecture used in this work.  Conv $W\times CI$ indicates a convolutional filter of size $W\times CI$, where $CI$ is the number of channels at the input. Feature size $L \times C$ indicates $C$ channels of feature vectors of size $L$. FC $544\times 2$ is a fully connected layer that takes features of size $544$  and transfers them to two dimensions. The input has only one channel as can be seen in the above network sketch.    \label{fig:network_sketch}}
\end{figure}

Fig.~\ref{fig:network_sketch} depicts the structure of the CNN we used. It consisted of four convolutional layers, all with stride $4$. Following deep learning terminology, this is equivalent to applying a regular convolution followed by sampling every fourth entry (discarding the rest). 
The first layer employed $40$ different convolution kernels of length $40$ (thus rendering the signal two dimensional, of dimensions $1527\times40$. The size of the second dimension is often referred to as the number of channels.). The second layer had $32$ kernels of size $23 \times 40$, meaning that each of the $32$ convolutions got its input from all $40$ channels, resulting in $32$ channels. The third and fourth layers had $32$ kernels of size $23 \times 32$.  The sizes of the kernels reflect a trade-off between minimizing the number of parameters we need to train, and maximizing the size of the receptive field of each 'neuron' (size of the input segment that affects it). A larger receptive field provides the ability to have more global decisions at the earlier layers of the network. The number of kernels learned at each layer should be large enough to include the variety of features extracted from the data at each layer. Yet it should not be too large, since from a certain point new features do not help the learning but rather add a computational burden and may lead to over-fitting.  We have taken all these considerations into account while designing the structure of the network.

Each convolution is followed by the ReLU non-linearity \citep{NaiHin2010}
and batch normalization (BN) \citep{IofSze2015}. The ReLU is an element-wise function that simply zeros all negative values in its input, without changing positive values, thus essentially performing a projection onto the positive orthant.  The role of the BN operation is to normalize the output of a given layer in the network such that its entries across different data samples have zero mean and unit variance. Yet it has trainable parameters that allow the network to deviate from this assumption, having other values for the mean and variance that enhance the overall performance of the network.  As BN ``stabilizes'' the data distribution at its output in the network, it improves the convergence during the training of the network. ReLU and BN  introduce the non-linearity to the network, which is an essential and crucial ingredient in each deep learning model. Both ReLU and BN are commonly used non-linearities in CNNs, where the first is referred to as the activation function of the network (trying to imitate the activation functions that appear in biological neurons) and the second is considered more as a regularization factor. The non-linearity components are fundamental in the design of any neural network.

At the end of the network, we employ a fully-connected layer (multiplication by a matrix) of size $544 \times 2$ followed by ReLU and a softmax operation \citep[i.e. normalized exponential, e.g.,][]{Bis2006}. The softmax, which is commonly used at the end of the network, is another non-linearity besides the ReLU and BN. It emphasizes the largest value in the input by applying an exponent element-wise followed by a division by the sum of the values. Thus, the output of the softmax component can be viewed as a probability distribution: each entry is in the range $[0,1]$ and the total sum is $1$. As a result, the output of the network consists of two values that indicate the probability (as estimated by the network) of the signal at the input of the network being either a noisy transit signal or pure (red) noise.  The detection is performed using one-hot-encoding: The output of the network is of the size of the number of classes to be detected (in our case noisy signal and pure noise) and each cell represents one of the classes. Thus, the ground truth labels of the output of the network  for noisy signal (or pure noise) are encoded by a vector that contains one at the entry corresponding to signal (or noise) and zero elsewhere.  

The loss function for training the network is the categorical cross entropy: Given a ground truth label $y$ and the output of the network $z$, it reads as
\begin{eqnarray}
-\sum_i y_i \log(z_i). 
\end{eqnarray}
The weights of the networks are updated such that this value is minimized on average over all the training examples. 
Notice that if the input belong to class $i = c$, then minimizing the above expression is equivalent to maximizing 
\begin{eqnarray}
 \log(z_c).
\end{eqnarray}
In order to update the weights of the network, we need the gradient of the loss function with respect to the network parameters. We calculate this gradient using the back-propagation algorithm \citep{Lecetal1998}, which is basically applying the chain rule iteratively throughout the network's layers.

Computing the gradient of the network based on all the data in the training set is prohibitively time consuming. Instead, we follow the common approach of stochastic gradient descent (SGD) with mini-batches. Unlike standard (batch) gradient descent that calculates the gradient of each step based on all the training examples together, SGD computes the gradient at each iteration based on only a small fraction of the data, which is drawn at random each time (without repetition). Once all the training examples have been used, a new training {\em epoch} starts, which re-use the data to continue updating the network's parameters. Our training process spanned $35$ epochs, during which the network converged. Training the network with few more epochs did not obtain significantly different performance.

Using SGD for the optimization poses two challenges: (i) SGD uses a step-size parameter, known also as the learning rate, which requires tuning; and (ii) its convergence given a fixed learning rate or a rate that decays in a pre-determined way is usually slow. Optimization literature, and the one related to deep earning in particular, proposes several strategies to face these issues. Due to computational complexity, using second-order methods is not feasible and only first order techniques are considered. In this work we rely on the ADAM optimizer \citep{KinBa2014}, which achieves faster convergence by adaptively setting the learning rate based on estimates of first and second moments of the gradients. We use the default parameters proposed with this method. 

The various components and techniques we used in the CNN and which we presented above, are basic tools of the trade in deep learning.  \citet{Gooetal2016} provides a clear and comprehensive introduction to deep learning, and explains those concepts more thoroughly. 

\section{Results} \label{sec:results}

\begin{figure}
\plotone{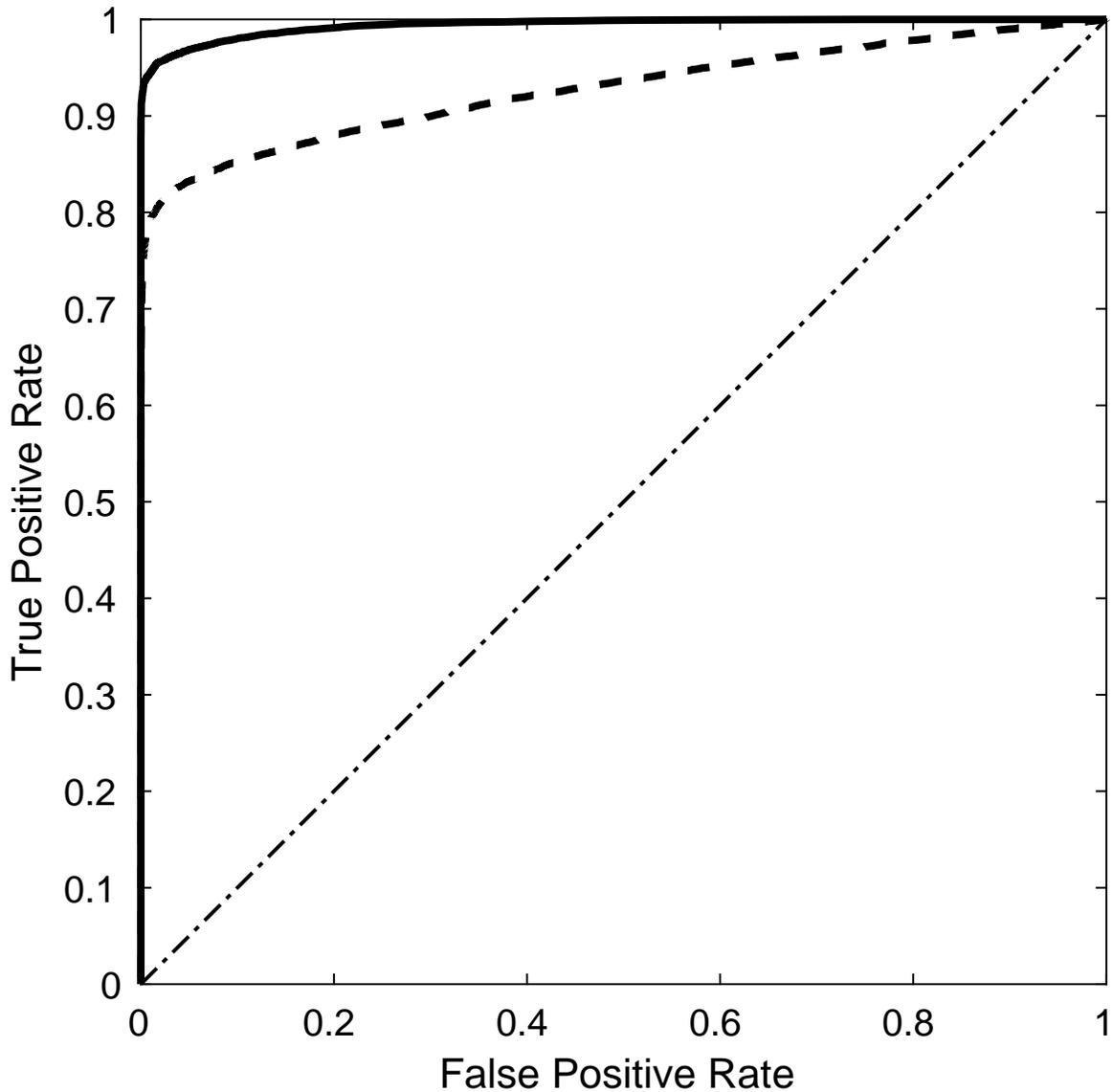}
\caption{Receiver Operating Characteristic (ROC) curve for the neural network and the dataset presented in this work. 
The dashed line represents the performance of the BLS preceded by a highpass filter.
The dot-dashed line is the so-called 'no-discrimination' line, corresponding to random guess.}
\label{fig:ROC}
\end{figure}

Detection was essentially determined by the terminal 'signal' value at the output layer, and whether it exceeded a predetermined threshold. In a similar fashion to traditional detection schemes, the value of that threshold determines the detection probability - both false detection and true detection. It is customary in detection theory to describe the behavior of a detection scheme using a Receiver Operating Characteristic (ROC) curve \citep[e.g.,][]{Faw2006}, which presents the true positive rate (TPR) as a function of the false positive rate (FPR). Fig.~\ref{fig:ROC} presents the ROC curve we got for our specific simulated dataset and the neural network described in Section \ref{sec:convnet} . 

For comparison with traditional approaches, we also ran the BLS algorithm \citep{Kovetal2002} on our simulated data. The transit period and duration ranges we used in BLS were exactly the ranges we used in the simulation. As we explained in Sec. \ref{sec:intro}, real fitting of a GP to each light curve is currently unfeasible. Instead, in order to have some reference performance, we used a simple approach of using a high pass filter, with a cut-off frequency of $1/6200\,\mathrm{min}^{-1}$, making sure we do not exclude the fundamental frequency of the longest period we seek. The test statistic we used for detection was the $Z$-score (which is essentially identical to the SDE of \citet{Kovetal2002}). For each threshold value of the $Z$-score we estimated the FPR using $83333$ pure noise light curves, and used different $16667$ transit-containing light curves for the TPR (in fact, it was the same division used for training and testing the neural network). The dashed ROC curve in Fig.~\ref{fig:ROC} represents the performance of the BLS.

In what follows we examine more closely the performance of the neural network in various cases. We fixed the FPR at $0.01$, which implied a TPR of $0.94$, or a miss rate of $0.06$. In absolute figures, we had a total of $166$ false positives, and $920$ misses.

\begin{figure}
\plotone{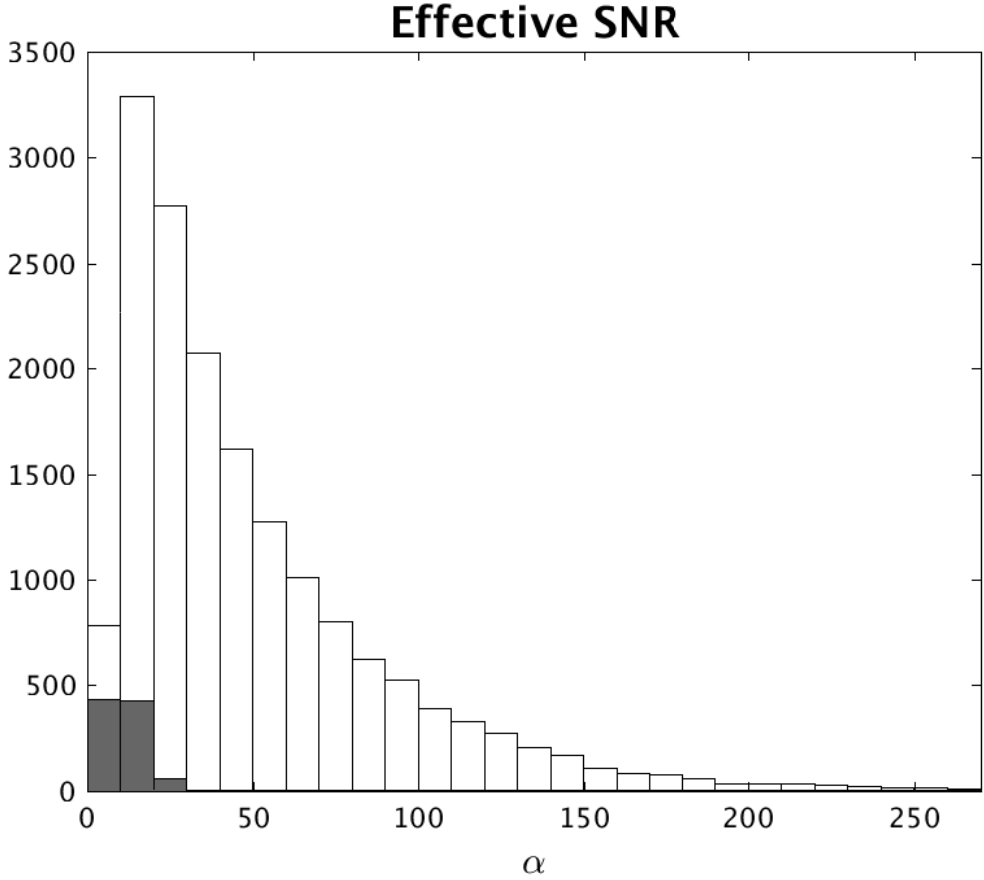}
\caption{A histogram of the effective SNR of the test light curves that include transit signals. The black bars represent the fraction of the non-detected transit signals (misses) \label{fig:hist_alpha}}
\end{figure}

\cite{Kovetal2002} defined an effective SNR which they denoted by $\alpha$:
\begin{equation}
\alpha = \frac{d}{\sigma}\sqrt{qn} \ ,
\end{equation}
where $q$ is the fractional transit duration, $n$ is the total number of samples, and $\sigma$ is the noise standard deviation. We calculated $\sigma$ directly from the simulated noise, including all the red- and white-noise contributions.  Fig.~\ref{fig:hist_alpha} presents the distribution of $\alpha$ for the test data (which include transits), while highlighting the distribution of the missed transits (as full histogram). It is obvious that the misses concentrate on the lowest end of the SNR range, as is intuitively expected.

\begin{figure}
\gridline{\fig{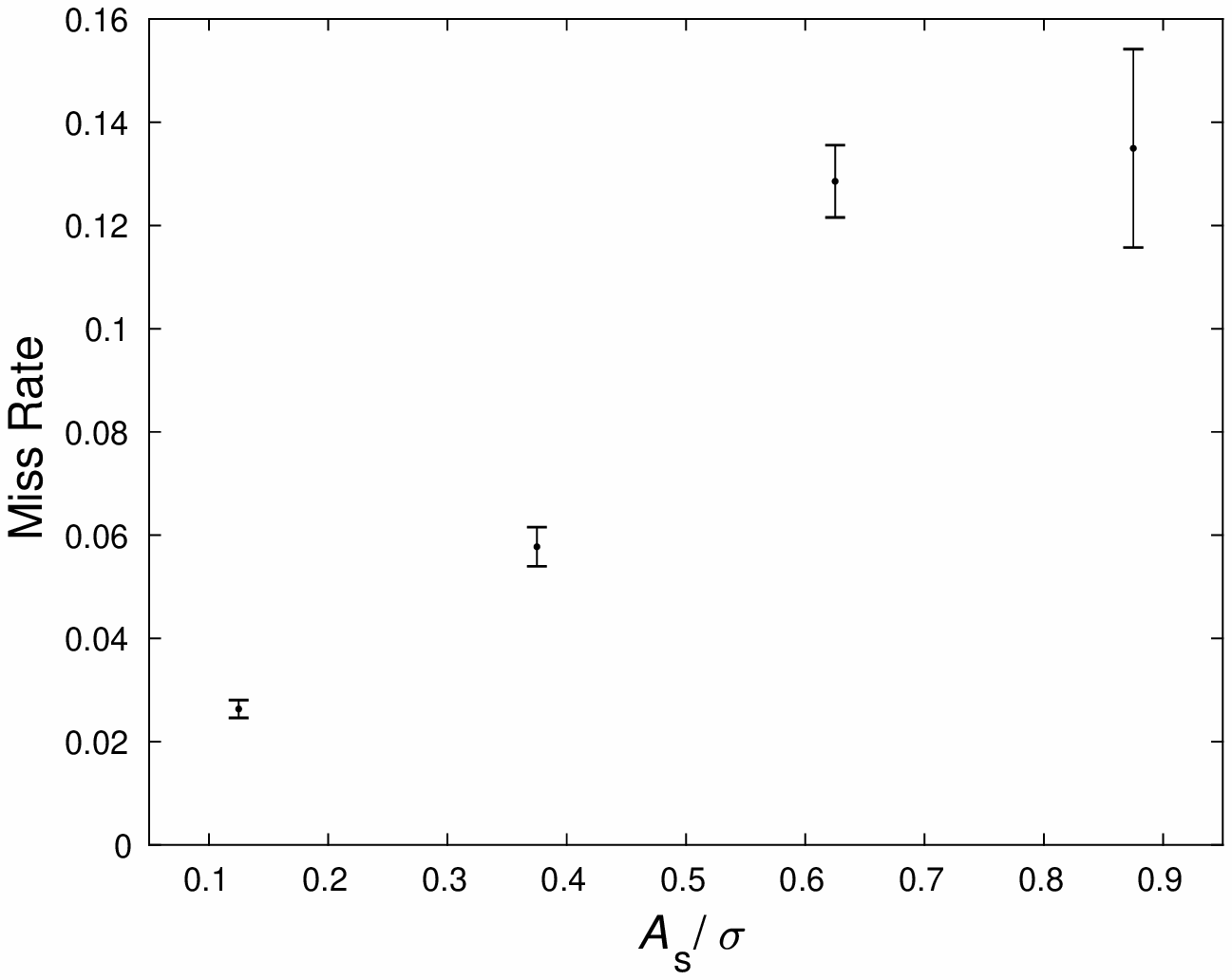}{0.3\textwidth}{(a)}
                    \fig{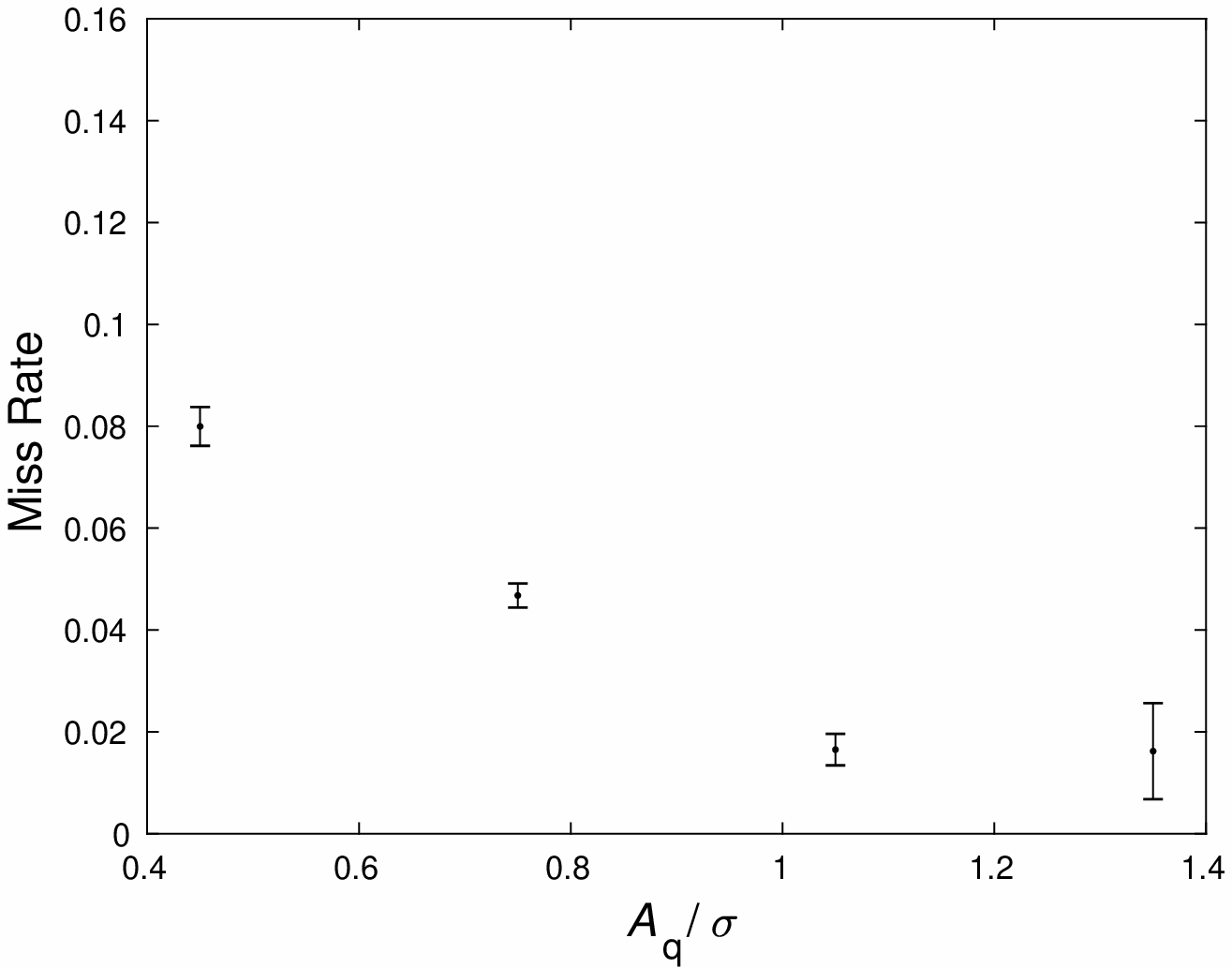}{0.3\textwidth}{(b)}
                    \fig{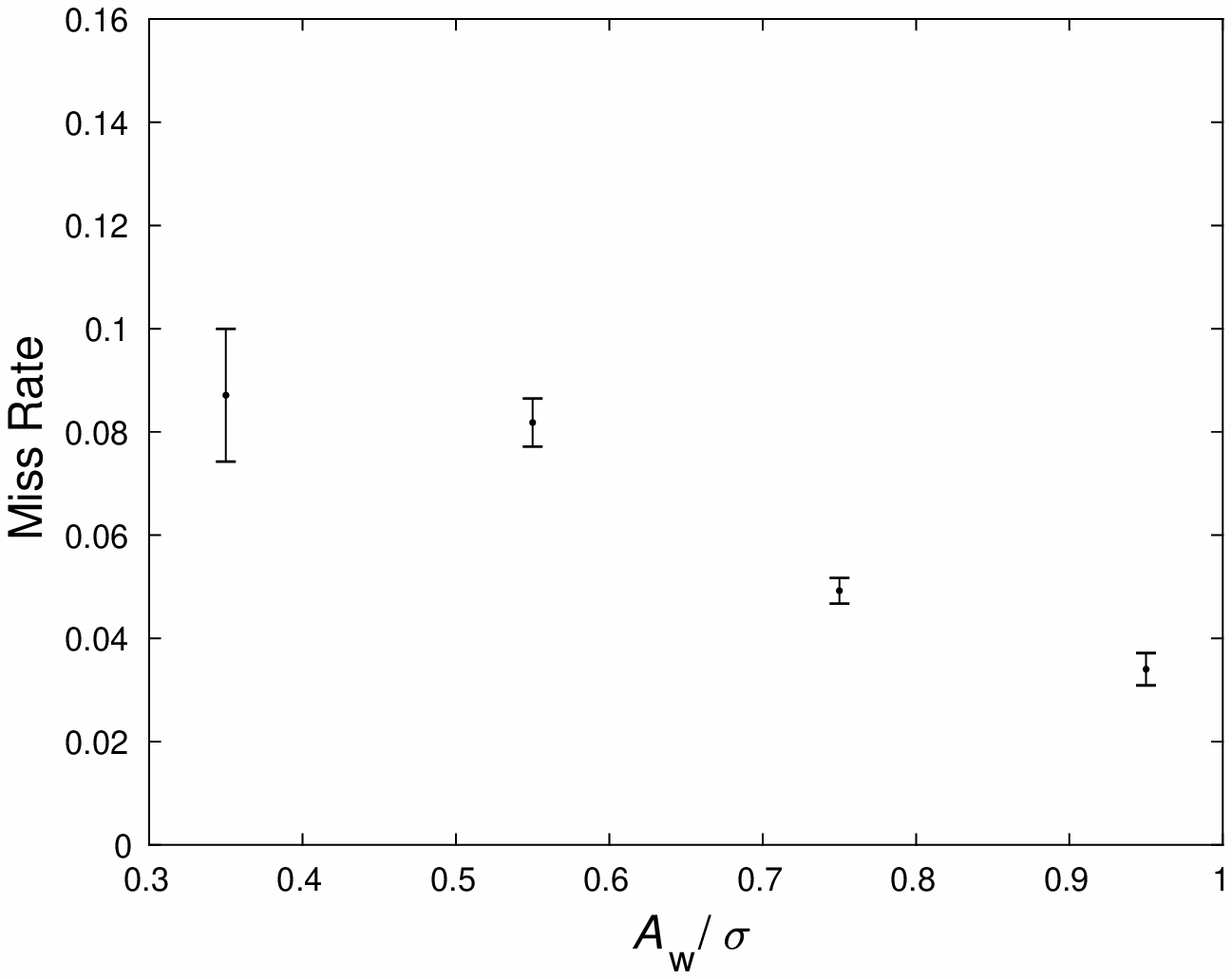}{0.3\textwidth}{(c)}}
\caption{The dependence of the false negative (miss) rate as a function of the relative contribution of each noise component --- squared-exponential (a), quasi-periodic (b) and white noise (c). \label{fig:fn_comp}}
\end{figure}

Fig.~\ref{fig:fn_comp} presents the dependence of the rate of misses as a function of the relative contribution of each noise component, represented by the ratio between the noise component amplitude and $\sigma$. Interestingly, it seems that the dominant factor affecting  the chances to miss a detection is the squared-exponential component.

\begin{figure}
\plotone{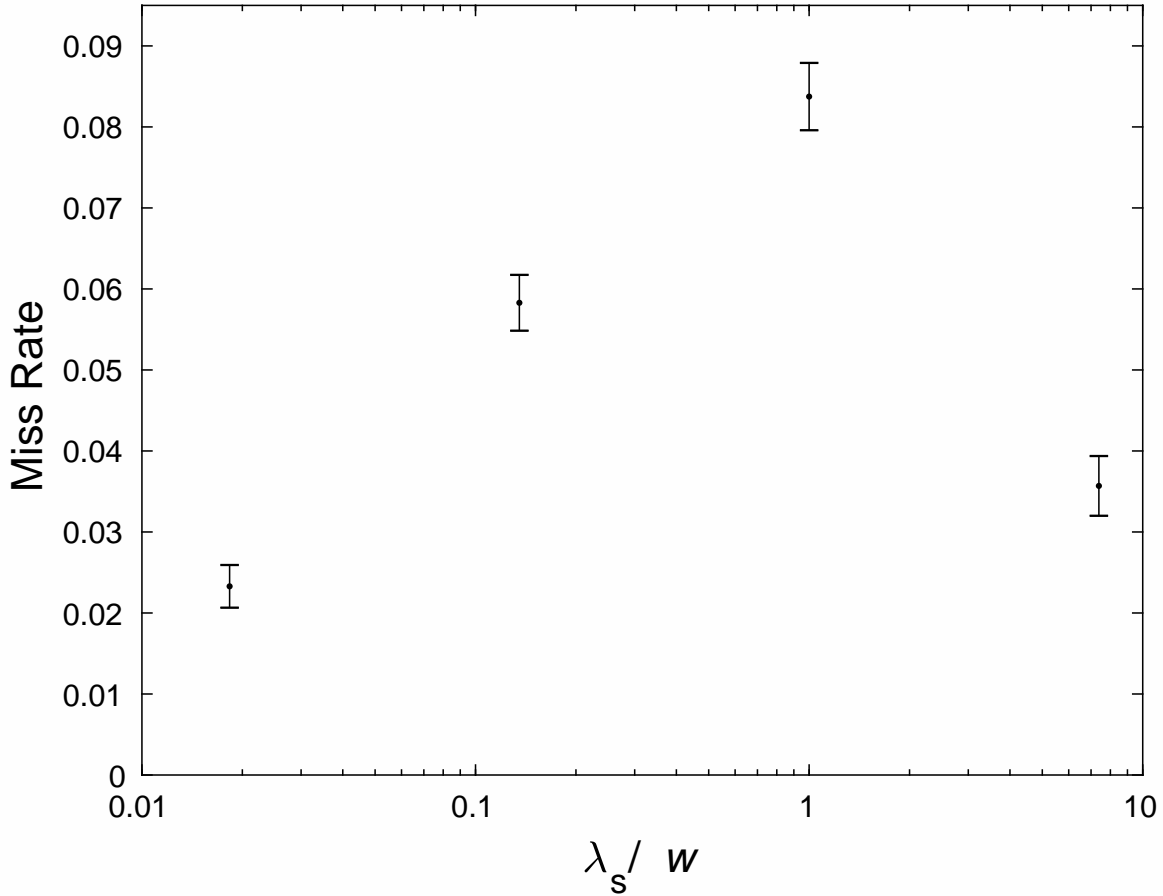}
\caption{The miss rate as a function of the ratio between the squared-exponential timescale $\lambda_\mathrm{s}$ and the transit duration $w$.
\label{fig:fn_lsw}}
\end{figure}

Taking a closer look at the effect of the squared-exponential noise, it is illuminating to examine the effect of $\lambda_s$ --- the timescale of this red noise component. Fig.~\ref{fig:fn_lsw} presents the dependence of the miss rate on the ratio of this timescale to the transit duration $w$. As can be expected, the effect on the miss rate is stronger when the two timescales are comparable, so that this red noise component might 'wash out' the transit events.

\begin{figure}
\plotone{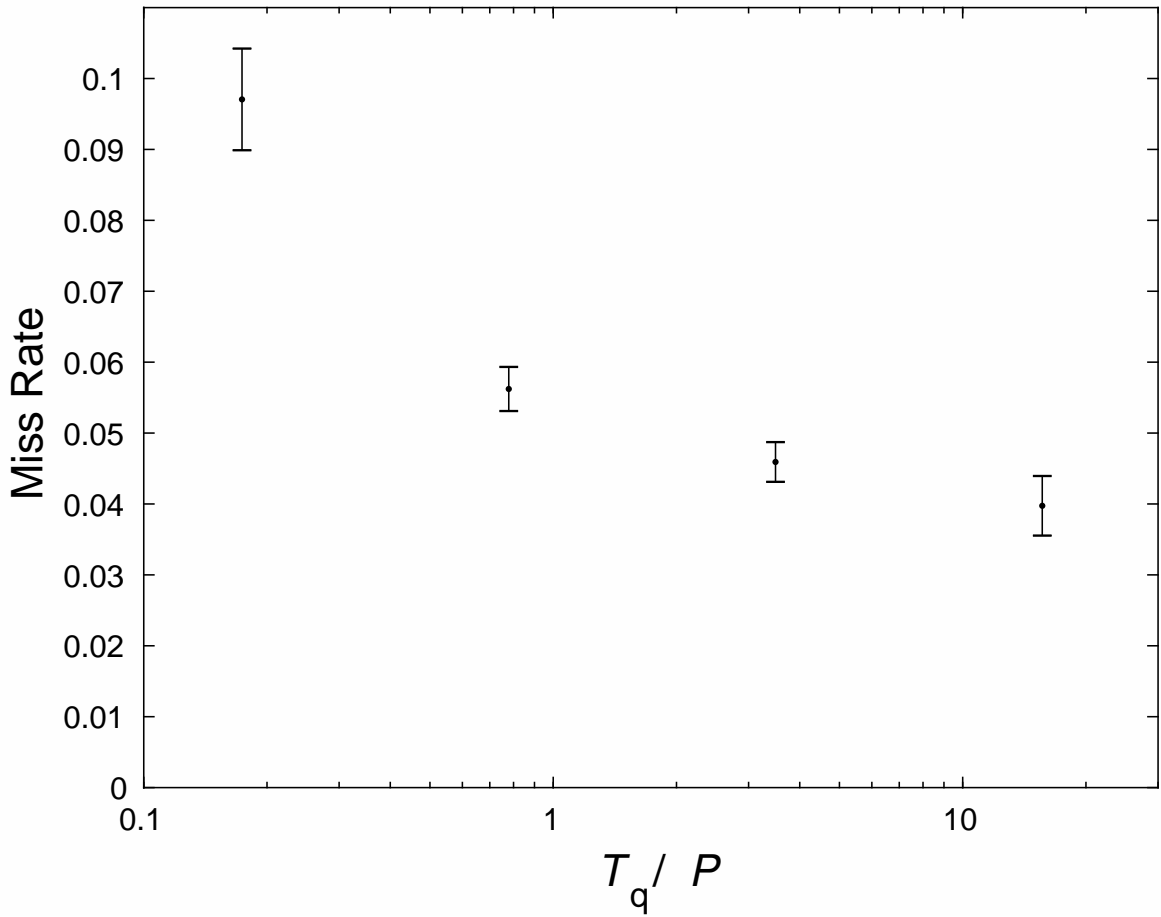}
\caption{Miss rate as a function of the ratio between the period of the quasi-periodic noise component ($T_\mathrm{q}$) and the transit period ($P$). \label{fig:fn_tqp}}
\end{figure}

In a similar way we checked the dependence of the miss rate on the period of the quasi-periodic noise component. Fig.~\ref{fig:fn_tqp} shows the dependence on the ratio between this period and the transit period. Here it seems that the main effect on the miss rate is when the noise period is shorter than the transit period, while noise periods that are longer constitute a much weaker disturbance to the detection.

\begin{figure}
\plotone{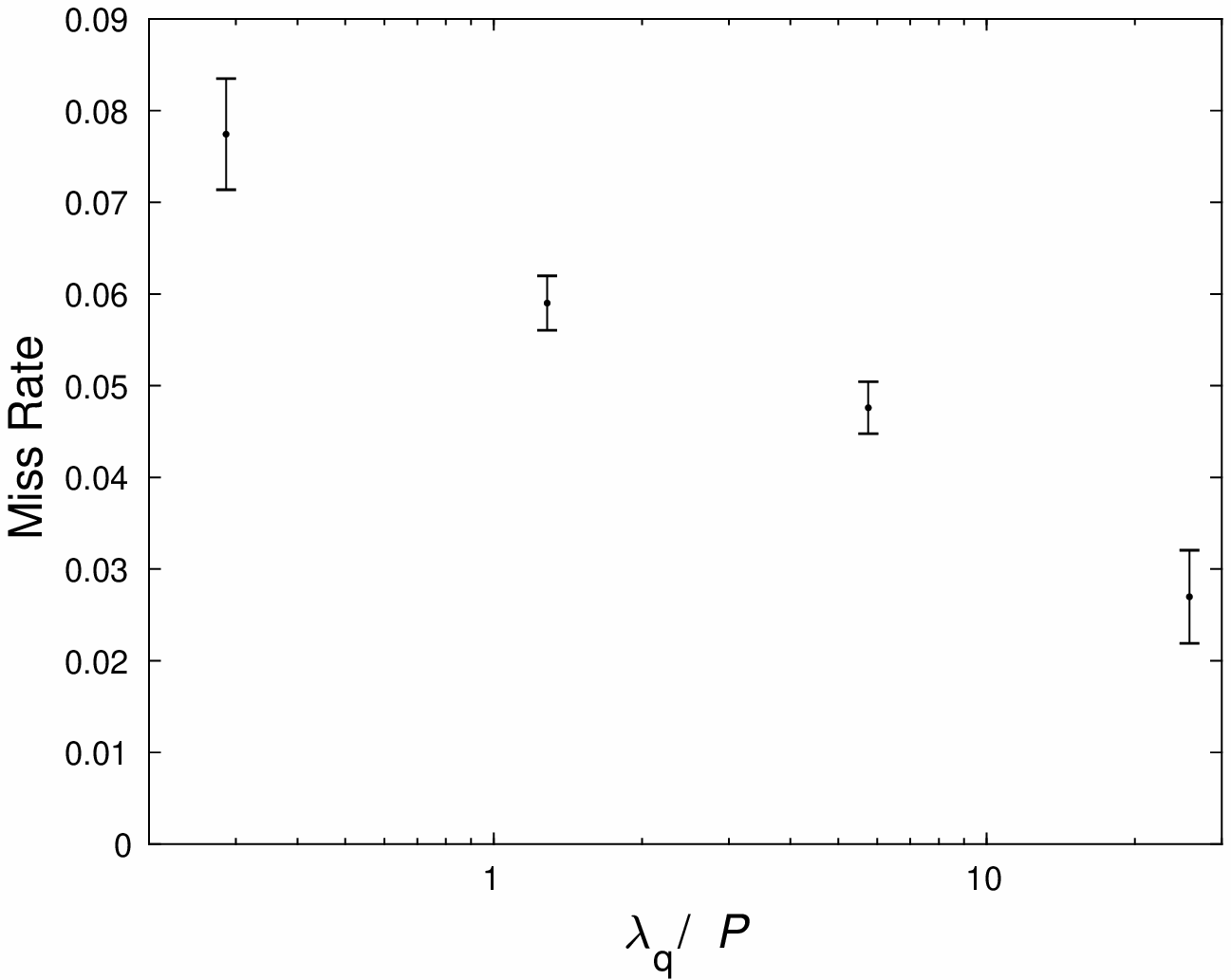}
\caption{Miss rate as a function of the ratio between the evolution timescale of the quasi-periodic noise component ($\lambda_\mathrm{q}$) and the transit period ($P$). \label{fig:fn_lqp}}

\end{figure}

The 'evolution timescale' of the quasi periodic kernel exhibits a similar behavior, as is shown in Fig.~\ref{fig:fn_lqp} --- evolution timescales that are shorter than the transit period, constitute a significant disturbance to detection.

\begin{figure}
\gridline{\fig{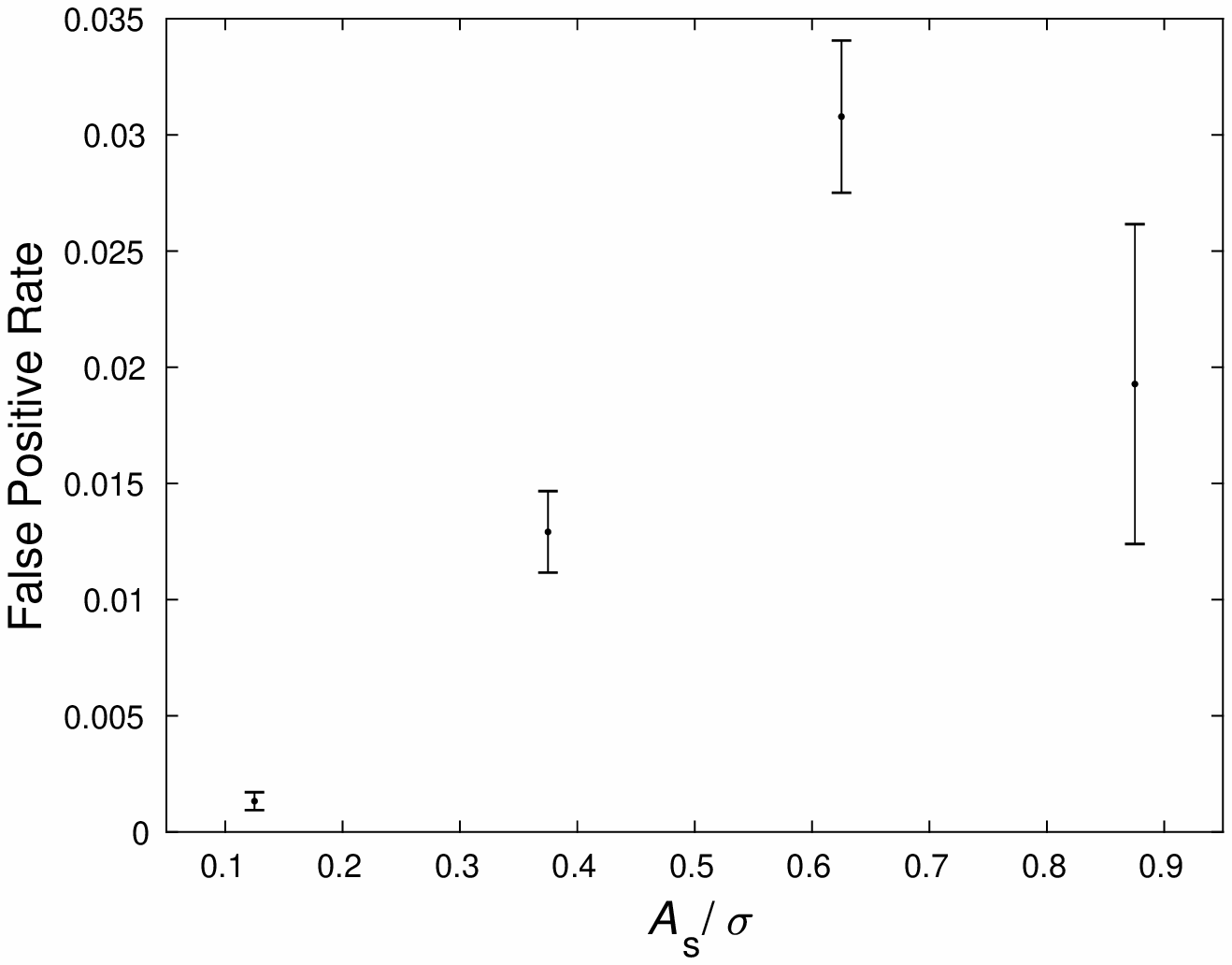}{0.3\textwidth}{(a)}
                    \fig{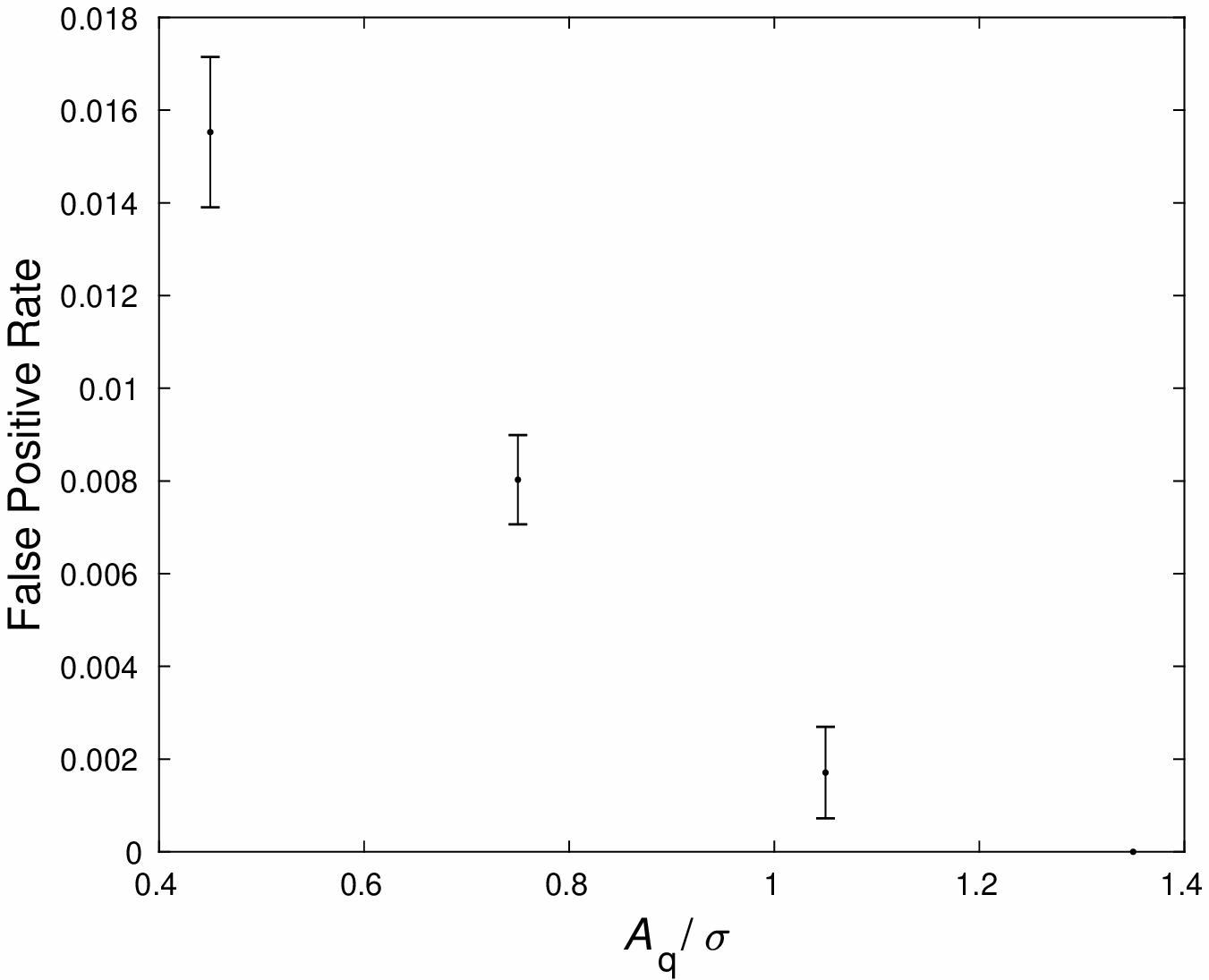}{0.3\textwidth}{(b)}
                    \fig{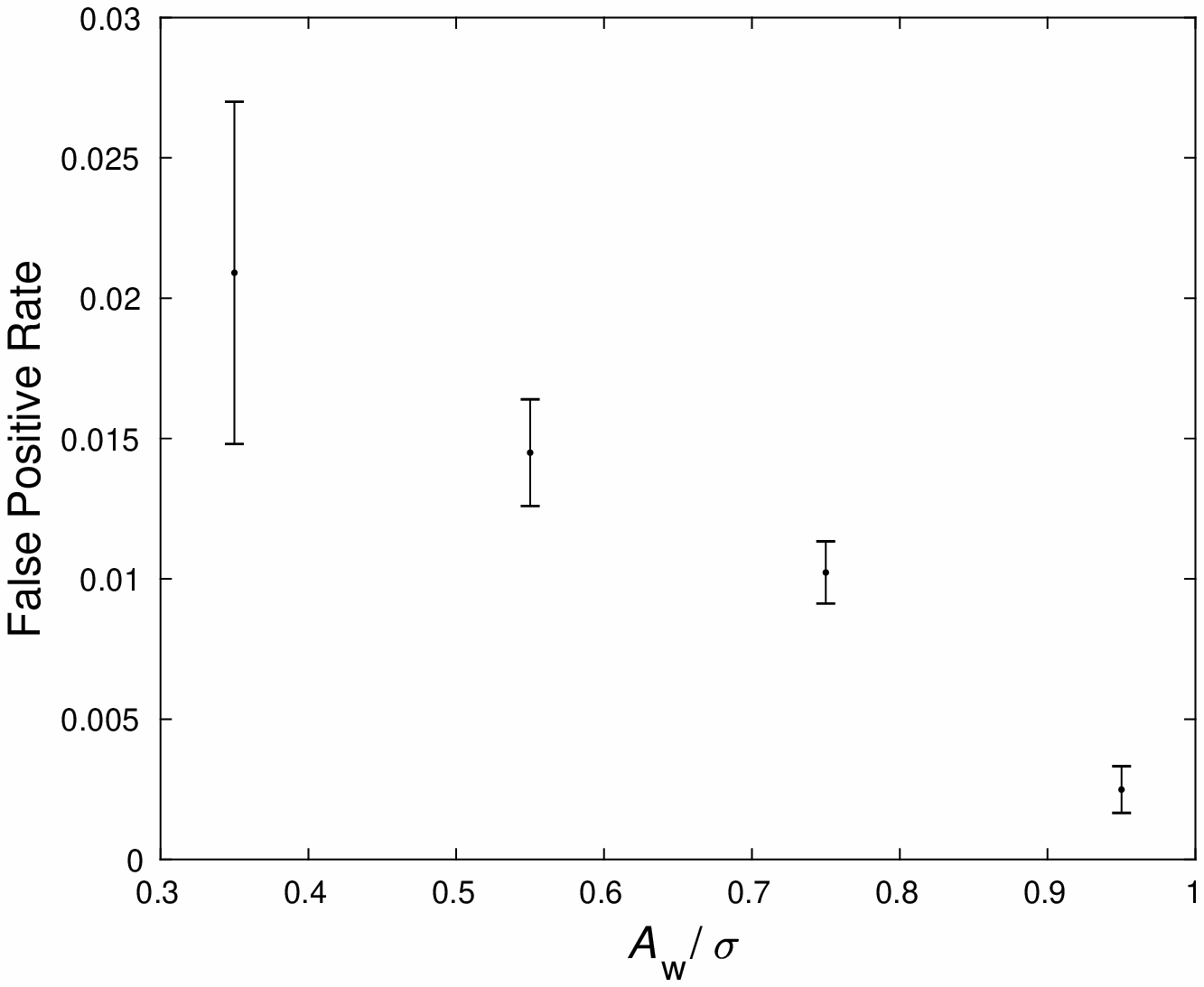}{0.3\textwidth}{(c)}}
\caption{The dependence of the false positive rate as a function of the relative contribution of each noise component --- squared-exponential (a), quasi-periodic (b) and white noise (c). \label{fig:fp_comp}}
\end{figure}

The importance of the squared-exponential noise component is also evident when we examine the false detections ('false positives'). Fig.~\ref{fig:fp_comp} shows the FPR as a function of the relative contribution of each noise component, same as in Fig.~\ref{fig:fn_comp}. Here again, the dominant factor is the amplitude of the squared-exponential component.

\begin{figure}
\plotone{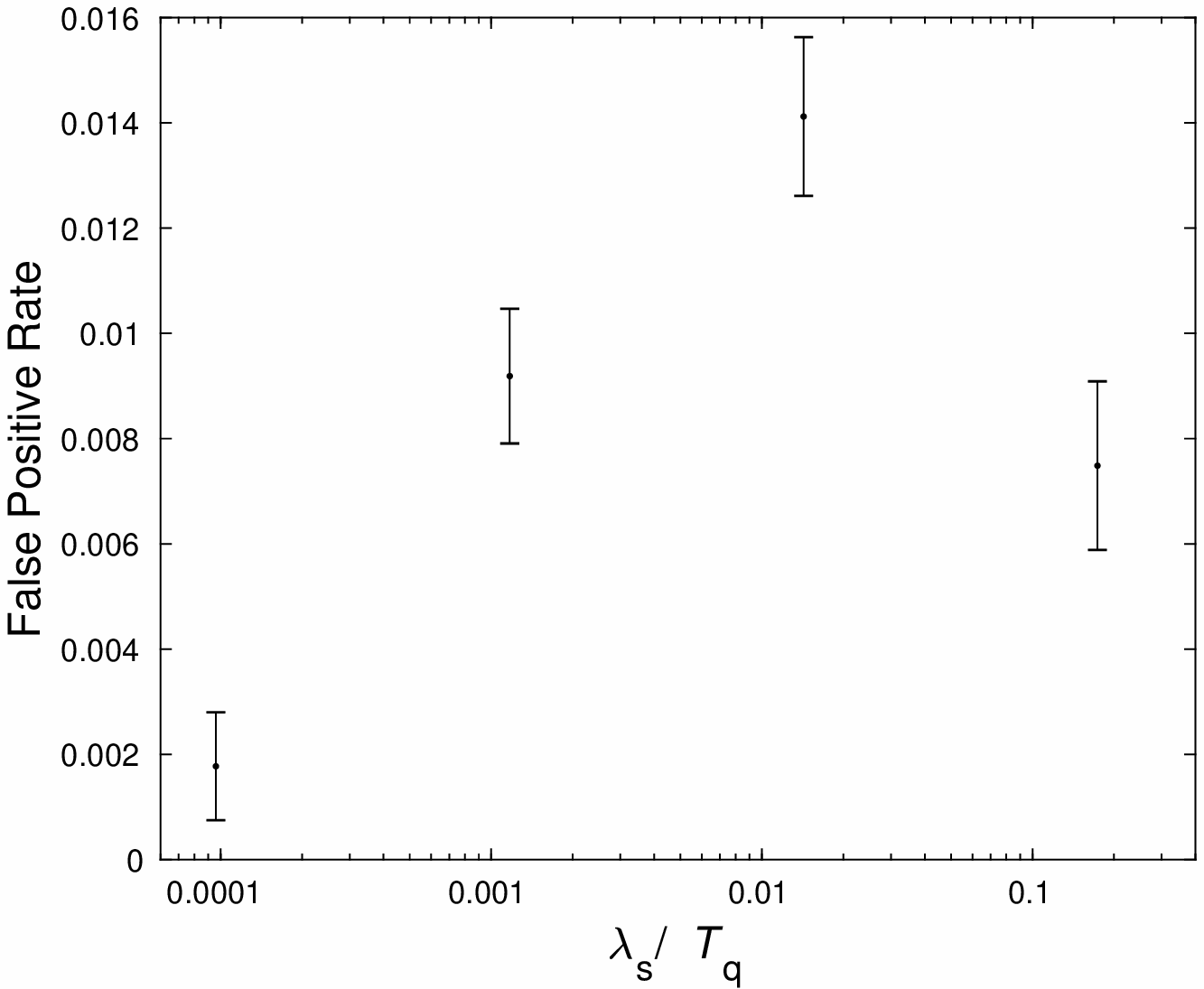}
\caption{False detection rate as a function of the ratio between the timescale of the squared-exponential noise component ($\lambda_\mathrm{s}$) and the period of the quasi-periodic component($T_\mathrm{q}$). \label{fig:fp_lstq}}
\end{figure}

In an attempt to characterize further the contribution of the various GP hyperparameters to false detections, the main effect we have found was that of the timescale of the squared-exponential component. Fig.~\ref{fig:fp_lstq} shows the dependence of the FPR on the ratio between this timescale and the period of the quasi-periodic noise component. the FPR is highest when this ratio is around a few percents. This might be related to the signals we trained the network to detect: the most common ratio between the transit width and the transit period in the training set is also around a few percents.

\begin{figure}
\gridline{\fig{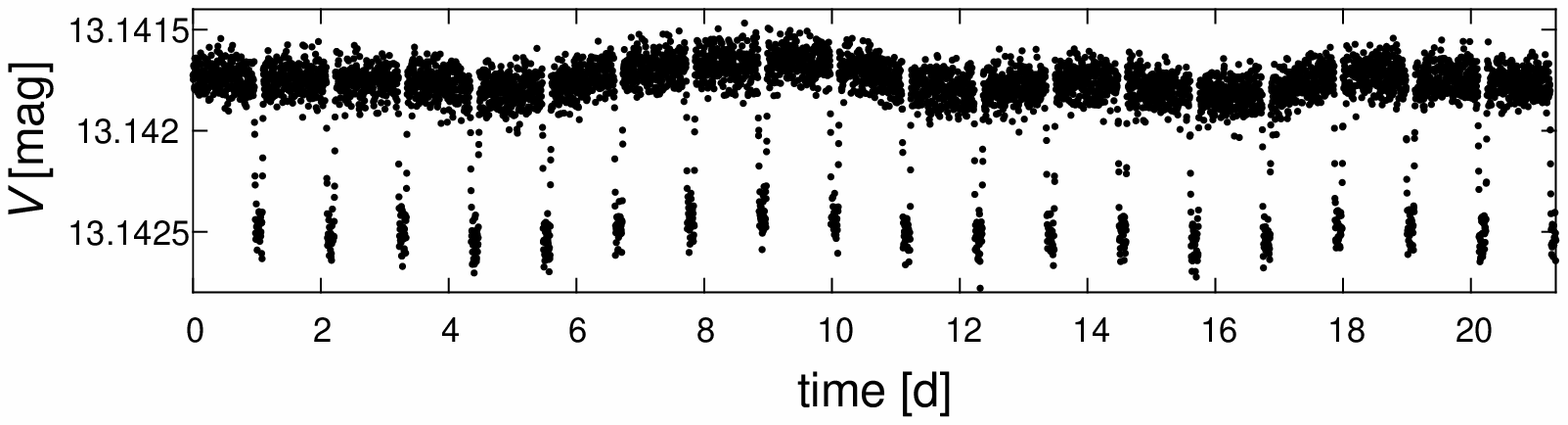}{1.0\textwidth}{(a)}}
\gridline{\fig{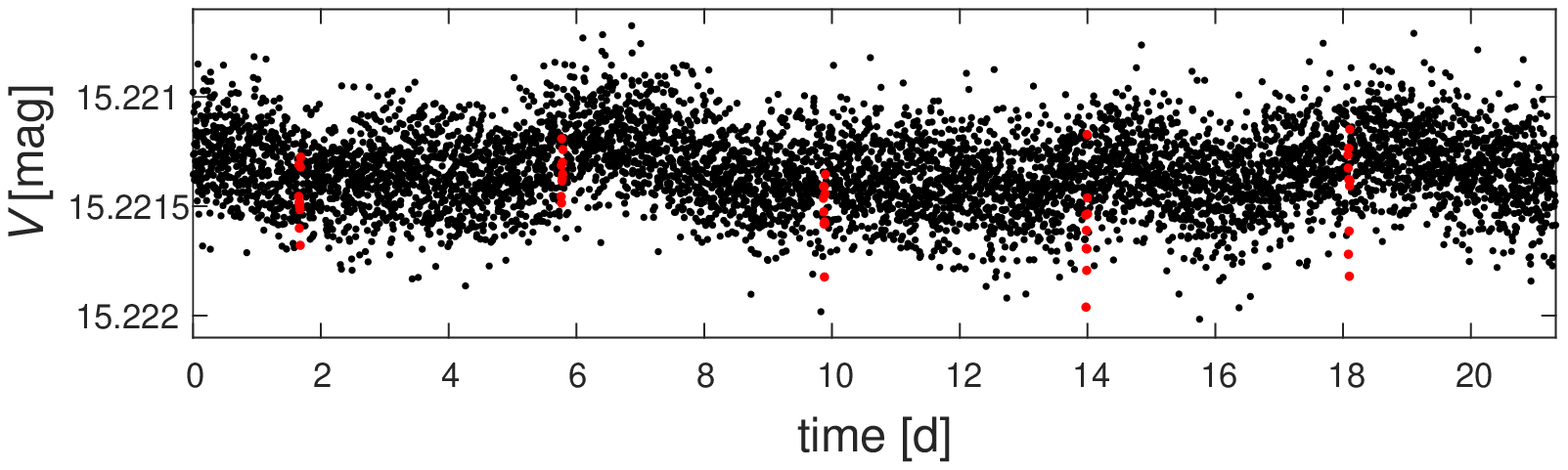}{1.0\textwidth}{(b)}}
\gridline{\fig{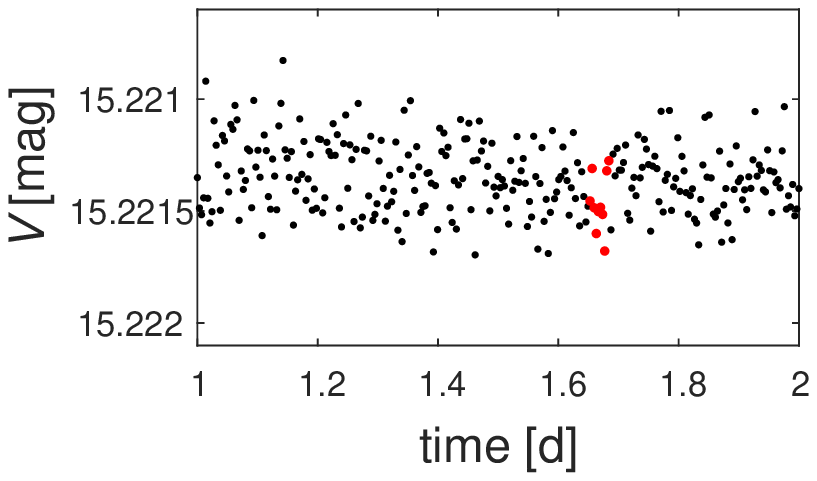}{0.3\textwidth}{(c)}
                     \fig{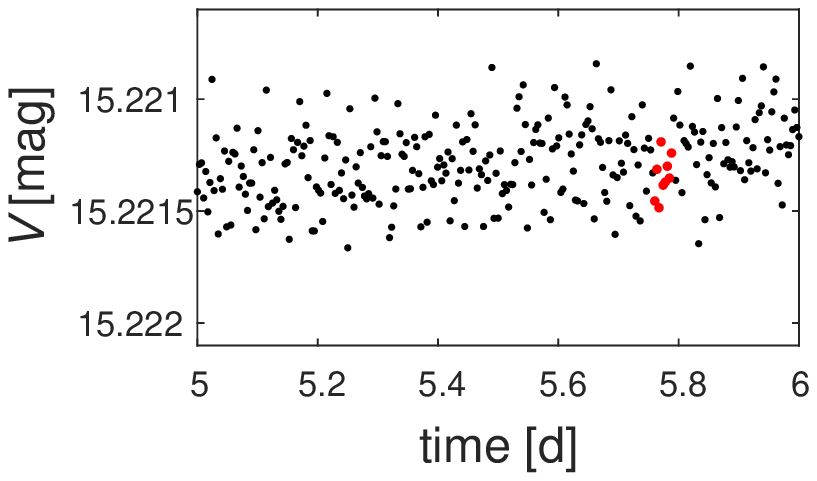}{0.3\textwidth}{(d)}
                     \fig{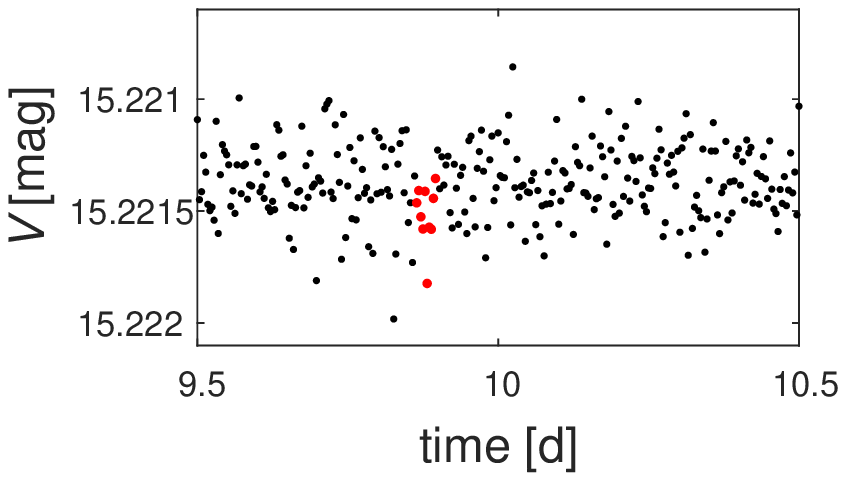}{0.3\textwidth}{(e)}}
\gridline{\fig{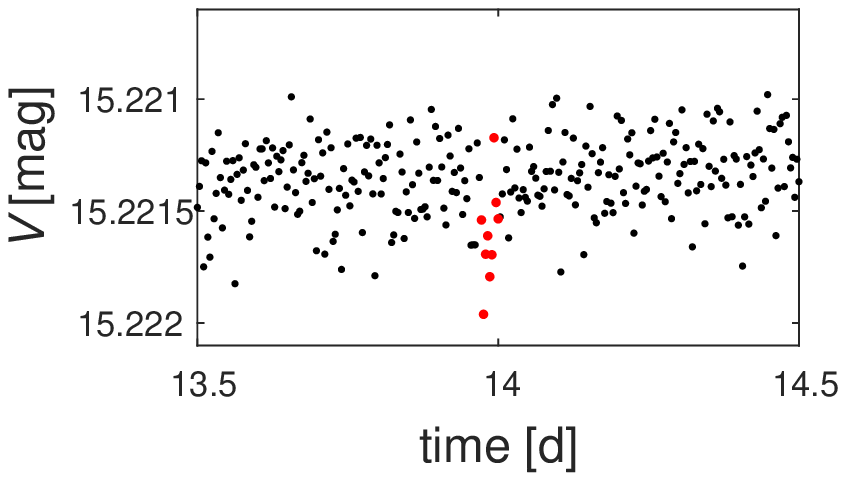}{0.3\textwidth}{(f)}
                    \fig{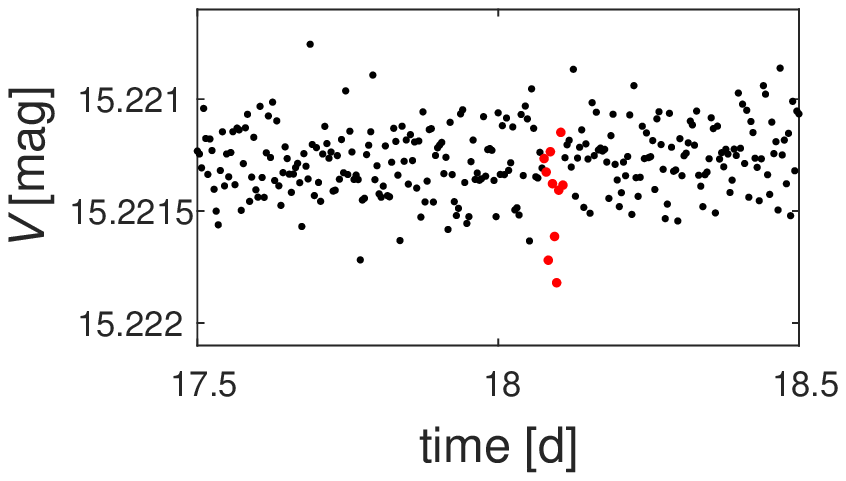}{0.3\textwidth}{(g)}}
\caption{Examples of detected transit signals ('true positives'). (a) A transit light curve with a high effective SNR of $\alpha=250.1$\,. (b) A transit light curve with a very low effective SNR of $\alpha=13.7$. (c) --- (g): Individual transit events of the light curve from panel (b).\label{fig:detections}}
\end{figure}

Finally, it is illuminating to examine the light curves of several representative cases. Fig.~\ref{fig:detections} shows two extreme examples of true detections by the neural network. The top panel (a),  demonstrates a case with a very high  effective SNR, $\alpha=250.1$. Obviously this is a very easy case to detect, and traditional techniques like the BLS would have probably detected it quite easily. There are many transit events, and they are much deeper than the noise by any standard. On the other hand, the bottom panel (b), shows a much more difficult case, with $\alpha=13.7$, on the lower end of the detectability histogram (see Fig.~\ref{fig:hist_alpha}). The individual transit events, with a depth of $0.133\,\mathrm{millimag}$, cannot be distinguished easily, so we highlighted them in red. Panels (c)---(g) zoom on the individual transit events, where one can see how difficult it is to spot them in the noise, e.g., the first two events (panels (c) and (d)) are not even fainter than the neighboring samples, and thus affecting the periodic nature assumed by the BLS and other methods. Nevertheless, the neural network did manage to detect the transit signal in this difficult case.

\begin{figure}
\gridline{\fig{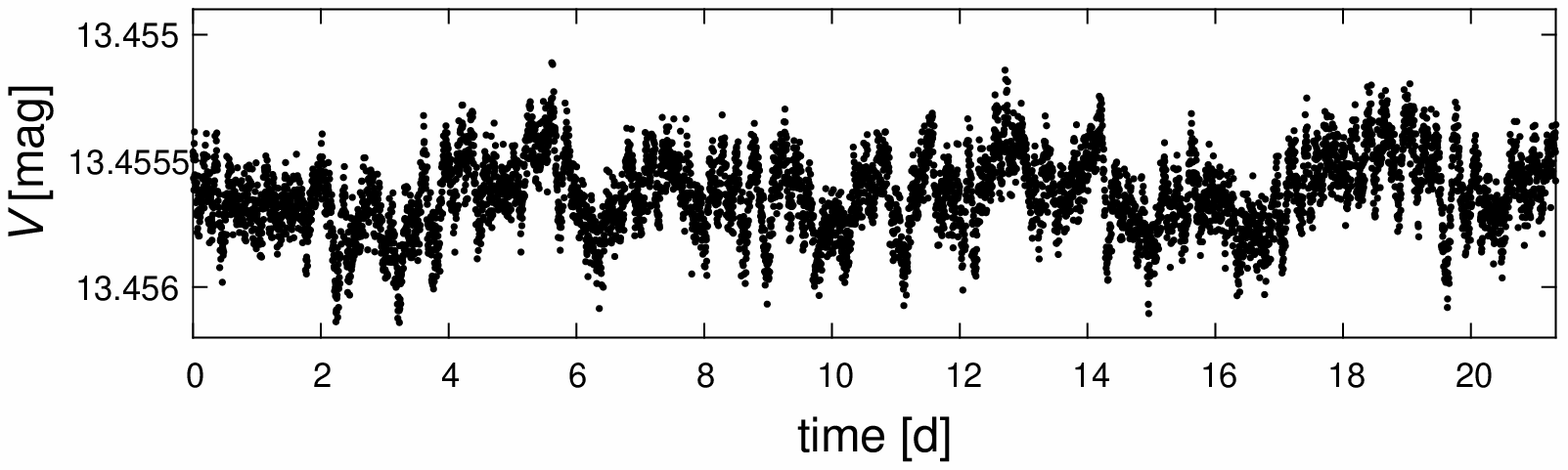}{1.0\textwidth}{(a)}}
\gridline{\fig{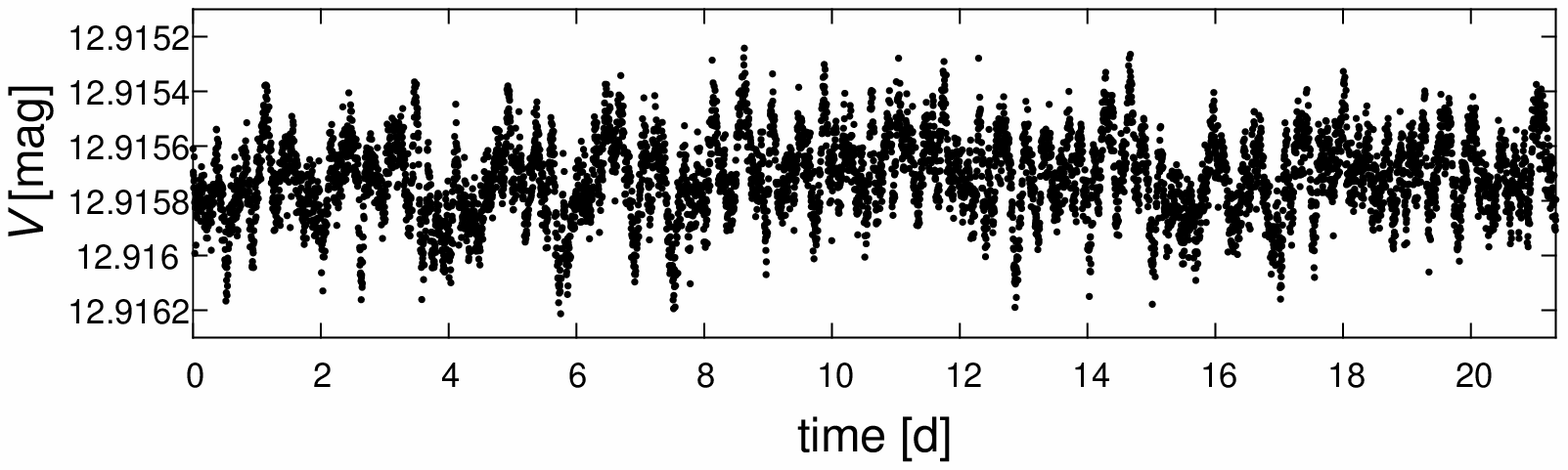}{1.0\textwidth}{(b)}}
\caption{Examples of light curves mistakenly tagged as transit detections. The two examples fit the trend in Figs. \ref{fig:fp_comp} and \ref{fig:fp_lstq}: (a) $A_s/\sigma=0.74$, $\lambda_s/T_q=0.022$, (b) $A_s/\sigma=0.83$, $\lambda_s/T_q=0.037$\,. \label{fig:fp} }
\end{figure}

In Fig.~\ref{fig:fp} we chose to present two examples of false detections, corresponding to the trend shown in Figs.~\ref{fig:fp_comp}~and~\ref{fig:fp_lstq}. It is apparent that those two light curves contain many events that may indeed resemble transits in a superficial glance, and thus might understandably lead to false detections.

\subsection{Outliers and Discontinuities} \label{subsec:outliers}

An important aspect of real-life light curves is the presence of artifacts caused by cosmic rays or abrupt changes related to operation of the observing apparatus. There are many methods to alleviate those effects. Even though these methods are usually quite satisfactory, those artifacts still constitute a source of confusion and false detections, and also mask out true signals. We tried to test our neural network against the inclusion of those artifacts as well. We therefore added to the simulated light curves outliers. Every sample in our light curves had a probability of $0.01$ to be affected by an outlier, that we implemented as an exponentially-distributed brightening, with an expectation value of $0.25$~millimag. Larger outliers are obviously easier to identify and remove, and we wanted to test our network in a more challenging setting. In addition to outliers we also added discontinuities. The discontinuity probability we implemented was $5\cdot10^{-4}$ per sample, and the value of the jump had a Gaussian distribution with a standard deviation of $0.5$~millimag. After each discontinuity we added an exponential decay to the previous level, with a timescale that was itself exponentially distributed with an expectation of $30$ minutes. Thus, some of the jumps had very long timescales that were effectively infinite, while others were so short that the discontinuity was effectively an outlier.

\begin{figure}
\plotone{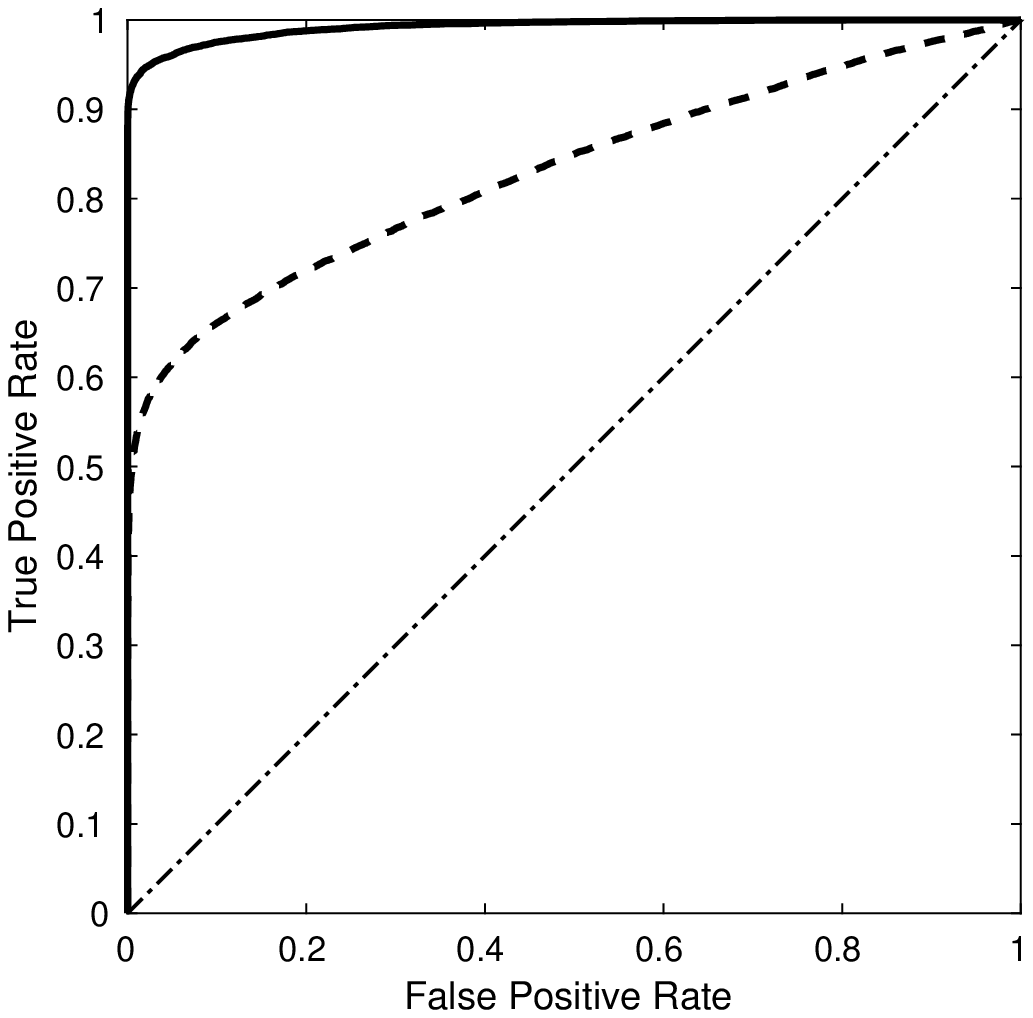}
\caption{Same as Fig.~\ref{fig:ROC}, for the dataset with the added artifacts of random outliers and discontinuities}
\label{fig:ROC_artifacts}
\end{figure}

We have repeated the process of training and testing the neural network on the new dataset, with the added artifacts, as well as the BLS and the high-pass filter. This time we preceded the high-pass filter with an outlier removal stage. This stage used calculating a moving median (on a $100$-sample window), and it rejected samples which deviated from this median more than three median absolute deviations. We also incorporated our knowledge that the outliers we considered were only brightenings, not dimmings. Fig.~\ref{fig:ROC_artifacts} presents the ROC we obtained for this new dataset. The performance of the neural network degraded by a marginal amount, while that of the BLS degraded much more significantly. 

\section{Discussion} \label{sec:discuss}

The results we presented in the previous section prove that a deep neural network can achieve satisfactory performance in detecting quite difficult transit signals in the presence of significant red noise. It is difficult to compare the performance to traditional techniques, as the standard techniques are optimized for white noise and require a preliminary stage of filtering out the red noise. There is currently no 'gold standard' that performs the two tasks, against which we can compare. 
We therefore compared our neural network performance to a na\"{i}ve combination of the BLS and a high-pass filter, designed to spare the relevant range of transit frequencies. The neural network performs much better!
In addition, one advantage of the approach we propose here is that it is not divided into two stages --- detrending and transit search. The neural network is fed the light curve, including the red noise, and it detects the transits without a detrending stage (which is probably performed implicitly and internally within the neural network mechanism). Furthermore, Section \ref{subsec:outliers} shows that the same neural network we used to detect transits in the presence of red noise, performs well also in the presence of outliers and jumps, after being appropriately trained. Thus, in principle, it can also save the need for the separate stages of outlier removal and discontinuities detection. All those abilities of the neural network are made possible by the highly nonlinear nature of the neural computation mechanism, which is totally different from the traditional methods, and opens a plethora of optional new capabilities.

As the results of Section \ref{sec:results} show, the detection performance of the neural network does behave in ways we would expect --- shallower transits, longer periods, lower effective SNR --- are all affecting the chances to detect the transits. In that respect, the results constitute a sanity check. One surprising result is that the squared-exponential component had more influence than the quasi-periodic component. However, when we turn to analyze real data, we will have to characterize the GPs in much more detail, with more components in the kernel, and this specific effect may turn out to be different in real life. 

In this context, of analyzing real data, we should emphasize the importance of the artificially generated training set. If our purpose would have been the detection of Jovian planets, we could, in principle,  skip the process of generating artificial light curves, and use light curves that seem not to contain transits. However, since we aim to detect planets that are difficult to detect, it seems imperative that the light curves we include in the training set as examples of light curves without transits, indeed be without transits. This can only be guaranteed by artificially generated light curves. 
In the future, we will investigate the question of the perfect composition of the training set: how much does the presence of transits in what we consider as transit-free light curves affect the performance? How much does deviation from the true noise model affect the performance? If it does, we will have to perform a meticulous mapping of the stellar variability space in terms of GP hyperparameters, to help us in compiling the training set. All these are future work research directions we intend to pursue.

One criticism against neural networks in this context is the fact that the trained neural network is a 'black box' in the sense that we are unable to trace the way it achieves the detection. This may indeed be  the case (although some 'reverse engineering' can be attempted), however, since we can characterize in significant detail the false positive rate, as we showed in the previous section, the detections are nevertheless useful scientifically. Furthermore, the feasibility study we presented here focuses on the most fundamental task in the process of detecting transiting exoplanets, namely, tagging light curves that contain transits. Usually, like in the case of the BLS algorithm, this task already performs also part of the task of characterization, and the detection tool also produces an estimate of the transit period. We chose to separate the two tasks, bearing in mind that probably the neural network maintains in its internal representation also an estimate of the period. This will be part of our future studies, where we will try to use deep learning to perform the related tasks of detrending and 'cleaning' the light curves (and perhaps characterizing the activity signal in passing), characterizing the transit signal, and identifying the individual transits for further analysis like identifying TTVs. It may very well turn out that the same neural network can be used for more than one task, but we chose to make minimal assumptions at this stage.

While we worked on the study we presented here, \citeauthor*{Peaetal2017} published a paper presenting similar ideas \citep{Peaetal2017}. There are several fundamental differences between the tests \citeauthor{Peaetal2017} present in  their paper and our test: their noise simulation does not use a GP approach, but only a quasi-periodicity heuristic combined with white noise. Furthermore, they draw the noise and transit simulation parameters from a discrete grid, as opposed to continuous distributions. This renders their training and test sets extremely unrealistic. In spite of those significant shortcomings, the study they present should still be considered a pioneering study that proposes the use of deep learning to detect terrestrial planets for the first time. Another recent paper that is worth mentioning here is the paper by \citet{ShaVan2017} who claim to have identified transiting planets using deep learning. However, the problem they tackled is fundamentally different from the problem of detecting transits. They have used a conventional approach to detect the transits and then applied neural networks to identify whether the events they had detected were caused by transiting planets or not. Nevertheless, their research is another milestone in the gradually growing recognition of deep learning potential within the exoplanet community.

In future works, applying the deep learning approach to real-life data will require considering more kinds of artifacts, e.g., sampling gaps. The transit signal will have to be more realistic, and allow complicating features, like multiple planets, {TTVs, and very long periods leading to very few transit events.

In the context of transiting planets, besides CoRoT \citep{Auvetal2009} and Kepler \citep{Boretal2010}, the scientific community is eagerly anticipating data to flow from NASA's TESS satellite, once it is launched and commissioned \citep{Ricetal2015}. We will use the data from TESS, in addition to those from CoRoT and Kepler, as the testbed on which deep-learning-based transit detection will be perfected and matured. The next major development will be the launch of PLATO by ESA \citep{Rauetal2014}. Applying deep learning on PLATO's data will probably result in a surge of terrestrial habitable planets. Obviously, implementing the ideas we present here on those real-life data will require scaling the networks to much longer light curves. To this end, High-Power Computing hardware will have to be used, both for training and testing. 

Deep learning is revolutionizing a growing number of areas in modern human life. It is now time for this development to revolutionize the search for extrasolar habitable planets as well.

\acknowledgments

We wish to thank the anonymous referee whose very useful comments helped us improve the paper considerably.
This research was supported by the Ministry of Science, Technology and Space, Israel.
R.G.\ acknowledges support by ERC-stg SPADE (grant No. 757497). 
This publication was aided by the international team led by J. Cabrera on ‘Researching the Diversity of Planetary Systems’ at ISSI (International Space Science Institute) in Bern, Switzerland.

\end{document}